\documentclass[12pt]{article}
\input epsf.sty
\newcommand{\ra}{\rangle}

\newcommand{\vv} {\bar v}
\newcommand{\uu} {\bar u}
\newcommand{\K}{{\rm K_1}}
\newcommand{\Kt}{{\rm \widetilde K_1}}

\newcommand{\wt}{\widetilde}

\newcommand{\be}{\begin{equation}}
\newcommand{\ee}{\end{equation}}
\newcommand{\ben}{\begin{eqnarray}\displaystyle}
\newcommand{\een}{\end{eqnarray}}
\newcommand{\refb}[1]{(\ref{#1})}
\newcommand{\p}{\partial}
\newcommand{\sectiono}[1]{\section{#1}\setcounter{equation}{0}}

\begin{document}
{}~
\hfill\vbox{\hbox{hep-th/0111281}\hbox{CTP-MIT-3213}
\hbox{CPGP-01/11-4}
\hbox{PUPT-2016}
}\break

\vskip .6cm

\centerline{\large \bf  Star Algebra Spectroscopy}

\medskip

\vspace*{4.0ex}

\centerline{\large \rm Leonardo Rastelli$^a$,
Ashoke Sen$^b$ and Barton Zwiebach$^c$}

\vspace*{4.0ex}

\centerline{\large \it ~$^a$Department of Physics }

\centerline{\large \it Princeton University, Princeton, NJ 08540,
USA}

\centerline{E-mail:
        rastelli@feynman.princeton.edu}

\vspace*{2ex}

\centerline{\large \it ~$^b$Harish-Chandra Research
Institute}

\centerline{\large \it  Chhatnag Road, Jhusi,
Allahabad 211019, INDIA}

\centerline {and}
\centerline{\large \it Department of Physics, Penn State University}

\centerline{\large \it University Park,
PA 16802, USA}

\centerline{E-mail: asen@thwgs.cern.ch, sen@mri.ernet.in}

\vspace*{2ex}

\centerline{\large \it $^c$Center for Theoretical Physics}

\centerline{\large \it
Massachussetts Institute of Technology,}

\centerline{\large \it Cambridge,
MA 02139, USA}

\centerline{E-mail: zwiebach@mitlns.mit.edu}

\vspace*{5.0ex}

\centerline{\bf Abstract} \bigskip

The  spectrum of the infinite dimensional 
Neumann  matrices $M^{11}$, $M^{12}$ and $M^{21}$ 
in the oscillator construction of the  three-string vertex  
determines key properties of the star product and of 
wedge and sliver states. We study the spectrum of eigenvalues and 
eigenvectors of these matrices using the derivation  
$K_1 = L_1 + L_{-1}$ of the star algebra, which 
defines a simple infinite matrix commuting with 
the Neumann matrices. 
By an exact calculation of the spectrum of $K_1$, and by
consideration of an operator generating wedge states, we are
able to find analytic expressions for the eigenvalues
and eigenvectors of the Neumann matrices and for the spectral 
density. The  spectrum of $M^{11}$  is continuous 
in the range $[-1/3, 0)$ with degenerate  twist even and 
twist odd eigenvectors for every eigenvalue  except for $-1/3$.

\vfill \eject

\baselineskip=16pt

\tableofcontents

\sectiono{Introduction and Summary} \label{sa0}

The star algebra of open string field theory (OSFT) \cite{OSFT} is an
infinite dimensional associative algebra on a 
space of open string 
fields.  While a precise and abstract 
mathematical characterization of this
algebra is not yet available -- mostly because it seems
unclear how to restrict the space of open string fields
to a suitable subspace where the desired axioms hold --
a description of the star product in terms of oscillator
expansions, or in terms of conformal field theory 
correlators, affords a 
concrete operational definition that can be used to 
star multiply certain string fields unambiguously.

Indeed, shortly after the construction of OSFT, explicit
oscillator representations of the star product in terms
of a three string vertex became available and explicit
tests of the axiomatic properties and of the formulation
were done \cite{gross-jevicki,gj2,cremmer,samuel}. 
This construction requires
the  choice
of a specific conformal field theory (CFT), 
and the most familiar one 
corresponds to the
background of a space-filling D25 brane. In this case the matter
part of the CFT is that of 26 free bosons, and the ghost part
of the CFT is that of the $(b,c)$ system.  In the matter part of
the oscillator construction, the three string vertex is built as an
exponential of a quadratic form in the matter oscillators. The
oscillators have mode labels extending over an infinite range, and string
state space labels, extending over three values. The matrices
$V^{rs}_{mn}$ defining  these quadratic forms, with $r,s = 1,2,3,$ and
$0\leq m,n \leq \infty $ go under the name of Neumann coefficients  and
they encode the concrete definition of star multiplication. 
In many cases 
it is convenient to treat the zero
modes  separately and regard $V^{rs}_{mn}$ for fixed $r,s$ as an
infinite  dimensional matrix with indices $m,n\ge 1$.
We
thus have nine infinite matrices. It turns out that out of these
nine matrices, cyclicity and symmetry properties imply that 
the information is contained  in three matrices $M^{11} = CV^{11}$,
$M^{12} = CV^{12}$ and $M^{21} = CV^{21}$, where $C$ is the twist
matrix $C_{mn} = (-1)^m \delta_{mn}$. These matrices formally
commute and as we will see 
they share eigenvectors. There are
additional relations which 
for a given eigenvector allow us 
to relate the 
eigenvalues of $M^{12}$ and $M^{21}$ to those of $M^{11}$. 
Therefore the study of the spectral properties of $M^{11}$ suffices.
For brevity we will simply call $M \equiv M^{11}$. 

\medskip
It has become clear over the last year that the
spectrum of $M$ controls several important properties  
of star products. For example, the normalization of star 
algebra projectors such as the sliver state, requires in the
matter sector determinant factors involving the matrix $M$
and the divergences in such factors are controlled by the spectrum
of $M$\cite{0008252,0102112}. Similarly, in vacuum string field 
theory \cite{0012251,0106010, 0111129}, the 
algebraic
prediction of ratios of tensions for D-branes of different
dimensionalities involves ratios of determinants of $M$ and 
an analogous matrix that includes oscillators with zero mode 
numbers \cite{0102112}.
Finally, formal properties, such as the commutation of the various
$M$ matrices can be rendered anomalous in the presence of inverses
of the factor $(1+ 3M)$ because of the presence of an eigenvalue
$\mu= -1/3$ of $M$ \cite{0108150,0111034,unpub, 0111069,0111153}.
Such manipulations are required in testing 
proposals for tachyon fluctuations in vacuum string field theory
\cite{0108150,0111034,0111153}.

The matrix $M$ not only describes the essence of the three
string vertex but is also intimately related to the so-called wedge
states \cite{0006240},  and to the sliver state. Indeed, if we star
multiply  two 
vacuum states we get a wedge state whose Neumann matrix is precisely
$M$. All matrices defining wedge states commute with $M$
and in fact have simple expressions in terms of $M$.  This
is also the case for the sliver.  Thus knowledge of the spectral
properties of $M$ allows us to understand wedge states quite
completely.

\medskip
In this paper we carry out a complete analysis of the spectrum of 
eigenvalues of $M$ and also find the corresponding eigenvectors. 
We begin by introducing the various conventions and definitions  in 
section \ref{sa1}. 
In section \ref{sa2} we establish the existence of 
an eigenvector 
of $M$ 
with eigenvalue $-1/3$ that makes the matrix $(1+3 M)$ singular. 
In  section \ref{sa3} we introduce a new matrix $\K$ 
with a continuous non-degenerate spectrum 
and find all its  eigenvalues and eigenvectors analytically. 
This matrix $\K$ is defined as the action of the star algebra
derivation $K_1 = L_1 + L_{-1}$ on the space of 
positively moded oscillators.
In section \ref{sa4} we show  that the 
matrix $\K$ commutes with $M$, $M^{12}$ and $M^{21}$. 
This together with the non-degeneracy of the $\K$ spectrum
implies that all the eigenvectors of $\K$ are eigenvectors
of $M, M^{12}$ and $M^{21}$. 
We also find the
precise relation between the eigenvalues  of $\K$ and $M$,
and give a functional interpretation of the eigenvalue  
equations. This interpretation extends the observation
of Moore and Taylor \cite{0111069} that the $C$-odd 
eigenvector of $M$ with eigenvalue $-1/3$ implies a flat
direction in the sliver functional. 

While it is in principle possible the $M$ has eigenvectors that
are not eigenvectors of $\K$ -- for example, a $C$ even eigenvector
of eigenvalue $(-1/3)$ -- our numerical experiments suggest that
we are not missing any 
piece of the spectrum. We thus
believe that the continuous spectrum of  $\K$ exhausts the 
continuous spectrum of $M$. The issue of the $C$ even eigenvector is
subtle since it is a vector that would be included
in addition to the continuous spectrum, and thus level expansion
experiments do not provide much insight. Our analysis could not
establish that this eigenvector belongs to the spectrum of $M$ and we
believe that it does not. Similar remarks apply to the matrices
$M^{12}$ and $M^{21}$.

Section \ref{sa5} is devoted to 
the study of the spectral density 
of $\K$ and $M$. For this purpose we consider the 
approximation of these matrices by $L\times L$ matrices.
We  find an explicit analytic expression for the 
density of eigenvectors in the large $L$ limit
and  compare it with numerical results finding reasonable
agreement.  We conclude in section \ref{sa9}  with some open questions
and remarks. 

\medskip  
\noindent
{\bf Brief Summary of Results.} Since the analysis of the  paper is
somewhat technical, we shall summarize  the main results here. 
The $*$-algebra derivation $K_1$ is represented on the space
of positively moded oscillators by a symmetric matrix 
$\K$ (equation \refb{kdata}). The spectrum of $\K$ 
\be
\K v^{(\kappa)} = \kappa v^{(\kappa)}\,,
\ee
exists for  $\kappa$ a real continuous 
parameter in the range $-\infty < \kappa < \infty$ and is
nondegenerate. For each $\kappa$, the eigenvector 
$v^{(\kappa)}$ has components
$v_n^{(\kappa)}$, with $n\geq 1$  given by the relation: 
\be \label{evnk}
\sum_{n=1}^\infty \, {1\over \sqrt n}\, v_n^{(\kappa)} z^n = {1\over 
\kappa} \Bigl(1 - \exp(-\kappa\tan^{-1} z) \Bigr)\, .
\ee
The derivation property of $\K$ ensures that
$[\K, M] = [\K , M^{12}] = [\K, M^{21}] = 0$. This together with 
the non-degeneracy of the $\K$ spectrum implies that the eigenvectors of
$\K$ are eigenvectors of $M, M^{12} $ and $M^{21}$. If we denote
by $\mu(\kappa), \mu^{12}(\kappa)$ and $\mu^{21}(\kappa)$ the
eigenvalues associated to $v^{(\kappa)}$ for $M, M^{12}$, and $M^{21}$ 
respectively, we find that
\ben \label{emuk}
\mu (\kappa) &=&\, -\, {1\over 1 + 2\cosh(\pi\kappa/2)} \, , 
\nonumber\\
\mu^{12}(\kappa) &=& -(1 + \exp (\pi\kappa/2) ) \, \mu (\kappa)  \\
\mu^{21}(\kappa) &=& -(1 + \exp (-\pi\kappa/2) ) 
\, \mu (\kappa)  \,.\nonumber
\een
Note that $\pm\kappa$ give the same value of $\mu$, thus each eigenvalue 
of $M$, except for  $\mu(0)=-1/3$, is doubly 
degenerate. 
Thus the spectrum of $M$ lies on the interval $[-1/3, 0)$ and is doubly
degenerate except at $-1/3$.\footnote{As mentioned before, a $C$-even
candidate at $-1/3$ exists, but it is not clear it properly
belongs to the spectrum of $M$.} It also
follows from the above that as
$\kappa\in (-\infty,
\infty)$,
$\mu^{12}$ grows monotonically from zero to one, while $\mu^{21}$ decreases
monotonically from one to zero. Thus, both $M^{12}$ and $M^{21}$ have
non-degenerate spectra in the interval $(0,1)$. 

The  eigenvectors $v^{(\kappa)}$ and $v^{(-\kappa)}$ 
are exchanged under the twist 
transformation. The degeneracy of $M$ allows us to 
introduce twist eigenstates
that are also $M$ eigenstates: 
\be \label{elcomb}
v_\pm^{(\kappa)} = {1\over 2} (v_n^{(-\kappa)} \mp v_n^{(\kappa)})\, .
\ee
For $\kappa=0$ we have a single $C$-odd eigenvector of $M$ with
eigenvalue $\mu(0) = -{1\over 3}$. The eigenvector is 
defined by taking the
right hand side of \refb{evnk} to be simply $\tan^{-1}(z)$.

If we approximate $\K$ by a matrix of size $L\times L$,
 the eigenvalues $\kappa$ of
$\K$ become discrete. For large $L$ the eigenvalues
approach a uniform distribution with density 
\be \label{eden}
\rho_\K^L(\kappa) = {1\over 2\pi} \ln L\, ,
\ee
where $\int_{\kappa_1}^{\kappa_2} \rho_\K^L(\kappa) d\kappa$ 
gives the number
of eigenvalues in the interval $(\kappa_1, \kappa_2)$. 
With the same finite approximation of $M$ the degeneracy between $C$-even
and $C$-odd eigenvectors is lifted.  Using
\refb{emuk} and
\refb{eden} one can easily find the  density of states in $\mu$ space to be 
\be
\label{mspecden}
\rho_{M}^L(\mu) = {2\over \pi^2}\,\,  {1\over |\mu|  
\sqrt{(1+ 3\mu) (1-\mu)}}\,  \ln L\, .
\ee

\sectiono{Notation and Definitions} \label{sa1}

The star product of two states $|A\ra$ and $|B\ra$ in the matter part of 
the conformal field 
theory is given by,\footnote{Our 
convention of $*$-product is the same
as that defined in ref.\cite{0105058}. The $V^{rs}$ appearing in 
eq.\refb{e9p} are transpose of the corresponding matrices given in the 
appendix A 
of ref.\cite{0102112}. With this the $*$-product defined here agrees with 
that given in ref.\cite{gross-jevicki}.}
\be \label{e8}  
|A*^mB\rangle_3 = ~_1\langle A| ~_2\langle B| V_3\rangle  \,,
\ee
where the three string vertex $|V_3\rangle$ is given by
\be \label{e9}
|V_3 \rangle = \int d^{26}p_{(1)}\, d^{26}p_{(2)}\, d^{26}p_{(3)} \,
\delta^{(26)}(p_{(1)} + p_{(2)}
+ p_{(3)})\,
\exp (-  E) \,
|0, p\rangle_{123} \,,
\ee
with
\be
\label{e9p}
E =  {1\over 2} \sum_{{r,s}\atop m, n \ge 1}
 \eta_{\mu\nu} a^{(r)\mu\dagger}_m V^{rs}_{mn}
a_n^{(s)\nu\dagger} +
\sum_{{r,s}\atop n\ge 1}
 \eta_{\mu\nu} p_{(r)}^{\mu} V^{rs}_{0n}
a_n^{(s)\nu\dagger}
+{1\over 2}\sum_r\eta_{\mu\nu} p_{(r)}^{\mu}
V^{rr}_{00} p_{(r)}^{\nu}  \, .
\ee
Here $a^{(r)\mu}_m$, $a^{(r)\mu\dagger}_m$ are non-zero mode matter
oscillators\footnote{In our notation \label{ff2} 
$i\sqrt{2} 
\p X^\mu(z) = \sqrt 2 p^\mu 
+ \sum_{n\ne 0} \sqrt{|n|} a_n^\mu z^{-n-1} = 
\sum_n 
\alpha_n^\mu z^{-n-1}$, and $\partial X^\mu(z) \partial X^\nu(w) \simeq 
-\eta^{\mu\nu} / 2(z-w)^2$.}
acting on the $r$-th string state
normalized so that
\be \label{e10}
[a^{(r)\mu}_m, a^{(s)\nu\dagger}_n] =
\eta^{\mu\nu}\, \delta_{mn} \, \delta^{rs},\qquad m,
n\ge 1\, .
\ee
$p_{(r)}$ is the 26-component momentum of the $r$-th string, and $|0,
p\rangle_{123}\equiv |p_{(1)}\rangle\otimes |p_{(2)}\rangle\otimes
|p_{(3)}\rangle$ is the
tensor product of the Fock vacuum of the three strings, annihilated by the
non-zero mode annihilation operators $a^{(r)\mu}_m$, and eigenstate of
the
momentum operator of the $r$th string with eigenvalue $p_{(r)}^{\mu}$.
$|p\rangle$ is normalized as
\be \label{enorm}
\langle p|p'\rangle = (2\pi)^{26}\delta^{26}(p+p')\, .
\ee
The
coefficients $V^{rs}_{mn}$ for $0\le m, n<\infty$ can be calculated by
standard 
methods~\cite{gross-jevicki,gj2,cremmer,samuel}. 

We define by $V^{rs}$ the matrices $V^{rs}_{mn}$ with $m,n\ge 1$, and by 
$C_{mn}$ the twist 
matrix 
$(-1)^m \delta_{mn}$. We also 
define:
\be \label{em11}  
M^{rs} = C V^{rs}\, .
\ee
Cyclic symmetry relates these matrices 
so that there are only three independent matrices $M^{11}$, $M^{12}$ and 
$M^{21}$. These matrices commute with each other 
and are real symmetric. 
Furthermore, we have the 
relations:
\be \label{em12m11}
M^{12}+M^{21} = 1 - M^{11}\, , \qquad M^{12} M^{21} 
= M^{11}(M^{11} -1) 
\ee
which allow us to determine the eigenvalues of $M^{12}$ and $M^{21}$ in 
terms of those of $M^{11}$.
Our main goal in this paper will be the determination of the eigenvectors 
and eigenvalues of the matrix $M^{11}$.
For convenience of notation, from now on 
we shall denote 
the matrices $M^{11}$ and $V^{11}$ by $M$ and $V$ 
respectively.

Finally, we note that in terms of the matrices $M^{rs}$ we can define 
projection operators\cite{0105058}:
\ben \label{edefrho}
\rho_1 &=& (1+T)^{-1} (1-M)^{-1} \Bigl( (M^{12} (1-TM) + T (M^{21})^2\Bigr)\, 
, \nonumber \\
\rho_2 &=& (1+T)^{-1} (1-M)^{-1} \Bigl( (M^{21} (1-TM) + T (M^{12})^2\Bigr)\,
, 
\een
where
\be \label{edeft}
T = (2M)^{-1} \Bigl(1+M - \sqrt{(1+3M)(1-M)}\Bigr)\, .
\ee
$\rho_1$ and $\rho_2$ can be shown to satisfy:
\be \label{erhoprop}
\rho_1^2=\rho_1, \qquad \rho_2^2=\rho_2, \qquad \rho_1\rho_2=0, \qquad 
\rho_1+\rho_2=1\, ,
\ee
and
\be \label{erhoprop2}
\rho_2 = C\rho_1 C\, .
\ee

\medskip   
We conclude this section by giving the explicit expressions for
the matrix $M$.  Following \cite{gross-jevicki,gj2} we have
\ben
\label{thematrix}
M_{mn} &=&  -{2\over 3}  {\sqrt{mn}\over m^2 - n^2} \Bigl (m A_m B_n - n A_n
B_m
\Bigr) \,, \quad m+n = ~\hbox{even},\,\,\,  m\not= n\,,  \nonumber \\
M_{mn} &=& \,\,\,\,0\,, \qquad\qquad\qquad\qquad\quad 
\qquad \qquad \quad m+n =
~\hbox{odd}\,, \\
M_{nn} &=& -{1\over 3}  \,\Bigl( 2 S(n) - 1 -(-1)^n A_n^2 \Bigr) \,,
\qquad S(n) = \sum_{k=0}^n  (-1)^k A_k^2 \,.\nonumber
\een 
In the above the coefficients $A$ and $B$ are defined as 
\be \label{ea1}
\bigg({1 + i x \over 1 - i x}\bigg)^{1/3} = \sum_{n\, even} A_n
x^n
+ i \sum_{n\, odd} A_n x^n\, , \quad
\bigg({1 + i x \over 1 - i x}\bigg)^{2/3} = \sum_{n\, even} B_n
x^n
+ i \sum_{n\, odd} B_n x^n\, .
\ee
The first few elements of the matrix are
\be
\label{mun}
M = \pmatrix{ -{5\over 27} & 0 & {32\over 243 \sqrt{3}} & 0 & - {416
\sqrt{5} \over 19683} & \cdots \cr
0& -{13\over 243} & 0 & {512\sqrt{2} \over 19683} & 0 & \cdots \cr
{32\over 243 \sqrt{3}}& 0 & - {893\over 19683} & 0 & {1504 \sqrt{5}
\over 59049 \sqrt{3}} & \cdots \cr
0& {512\sqrt{2} \over 19683} & 0 & - {5125\over 177147} & 0 & \cdots \cr
- {416 \sqrt{5} \over 19683}& 0 & {1504 \sqrt{5}
\over 59049 \sqrt{3}} & 0 & - {41165\over 1594323} & \cdots \cr
\vdots & \vdots &\vdots &\vdots &\vdots &\vdots  }
\ee
The CFT method furnishes an integral expression for the
elements of $M^{11}$. For this purpose we note  
the general formula\cite{lpp}
\ben  \label{evmn}
M_{mn} = {(-1)^{m+1}\over \sqrt{mn}}
\oint_0{dw\over 2\pi i}
\oint_0{dz\over 2\pi i}
\,{1\over z^m w^n} {f'(z)
f' (w)\over (f(z) - f(w))^2}\,,
\een
where, for the three string vertex, 
\be \label{efz}  
f(z) =\Bigl( {1+iz\over 1-iz}
\Bigr)^{2/3}.
\ee 
Both $w$ and $z$ integration contours are circles around the origin, 
with the $w$ contour lying outside the $z$ contour, and both contours 
lying inside the unit circle.

\sectiono{The $-1/3$ Eigenvector of $M$} \label{sa2} 

In this section we shall show that the matrix $M$ has an eigenvector 
$v^-$ with eigenvalue $-1/3$.
This eigenvector 
turns out to be $C$ odd (thus the
label), and  equivalently  $v^-$ satisfies $V v^- =
{1\over 3} v^-$, since $V = CM$.  We will establish this
result using  the conformal field theory representation of the vertex 
$V_{mn}=(-1)^m M_{mn}$.  
The required  expression was given in the previous subsection. 
Using integration by parts in $w$, and \refb{efz}, 
eq.\refb{evmn} 
can be turned into
\ben
\label{vsimp}
V_{mn} = -{4i\over 3} \sqrt{n\over m}
\oint {dw\over 2\pi i}{1\over w^{n+1}} 
\oint {dz\over 2\pi i} {1\over z^m} {1\over 1+ z^2}
{f(z)\over
f(z) - f(w)}\,.
\een
Numerical work suggested that the eigenvector $v^-$ was of the form
\ben
\label{eigen33}
v_n^-=  (-1)^{(n-1)/2} {1 \over \sqrt{n}} \,, \quad  
\hbox{for} \, \, n~
\hbox{odd}, \qquad v_n^- = 0 \,, \quad 
\hbox{for} \, \, n ~\hbox{even}\,.
\een
We shall show that $v^-$ defined in eq.\refb{eigen33} is indeed an 
eigenvector of $V_{mn}$ with eigenvalue $1/3$.
For regulation purposes and to understand
what residues are to be picked up, take a real
number $a$ slightly bigger than one and write
\be
\label{regv}
v_n^- = {a^{-n-1}\over 2\sqrt n} \{ (i)^{n-1} + (-i)^{n-1}\}\, . 
\ee
We understand that 
the limit $a\to 1^+$ is
to be taken.  Using equations \refb{vsimp} and \refb{regv} we
get  
\be \label{eneweq}
\sum_n V_{mn} v_n^-
= -{4i\over 3} {1\over \sqrt{ m}} \oint {dw\over 2\pi i}
\oint {dz\over 2\pi i} {1\over z^m} {1\over 1+ z^2}
{1\over (1+ a^2 w^2)}
{f(z)\over f(z) - f(w)} \, .
\ee
In order to be able to carry out the sum over $n$ to arrive at the above 
equation, we must have $|w|>a^{-1}$. 
Thus the $w$ integral 
picks up contribution from the poles at $w= \pm i/a$ and $w=z$. 
After this we can 
set $a=1$.
In this case only the $w=z$ and the $w= i$ poles
contribute since  $f(-i) = \infty$. Their contributions give
\ben
\sum_n V_{mn} v_n^-
= \Bigl( 1 - {2\over 3} \Bigr) {1\over \sqrt{ m}}
\oint {dz\over 2\pi i} {1\over z^m} {1\over 1+ z^2}
= {1\over 3}\, v_m^-\, .
\een
This establishes the claim.

\sectiono{$\K$ and its Eigenvectors} \label{sa3} 

In this section we shall introduce a matrix $\K$ representing
the action of the star-algebra derivation $K_1 = L_1 + L_{-1}$.
We shall be able to find explicit forms for the 
eigenvectors and eigenvalues of $\K$. In particular we shall see
that $\K$ has a non-degenerate continuous spectrum. In the next
section we shall show that $\K$ and $M$ commute, and thus the
eigenvectors of $\K$ are eigenvectors of $M$. Further analysis
will reveal the relation between the eigenvalues.

\medskip
We begin our analysis by recalling that 
the operator
\ben
K_1 = L_1 + L_{-1}
\een
is a derivation of the
star algebra \cite{wittensupersft, 0006240}.
We use its
action on positively moded oscillators $a_n$ and 
\be 
\label{aalpha}
\alpha_n\equiv \sqrt{n} 
a_n
\ee
 with $n\geq 1$ 
to define  matrices $\K$ and  $\Kt$
\ben
\label{intm} 
&&  [K_1, v \cdot a] \equiv (\K v) \cdot a - 
\sqrt 2 v_1 p  
\nonumber \\
&&  [K_1, w \cdot \alpha] \equiv (\Kt w) \cdot \alpha - w_1 \alpha_0\, .
\een
Here we have introduced the vector notation
\be
x \cdot a \equiv \sum_{n=1}^\infty x_n a_n,  \quad
y \cdot \alpha \equiv \sum_{n=1}^\infty y_n \alpha_n \,,
\ee
and suppressed the Lorentz indices.
Identifying $v\cdot a$ to $w\cdot\alpha$ and using eq.\refb{aalpha}
we get, 
\be \label{vw}
v_n = \sqrt n \, w_n\, .
\ee
{}From (\ref{aalpha}), we have the relation
\be \label{kkt}
(\Kt)_{mn} = \sqrt{n\over m} (\K)_{mn}\, .
\ee
Using the standard commutators:  
\be \label{ecomm}
[L_m, \alpha^{\mu}_n] = - n\, \alpha^\mu_{m+n}\, , 
\ee
we see that
the matrix $\K$ is symmetric, the diagonal elements are zero,
and the only non-vanishing entries are one step away from the
diagonal
\ben 
\label{kdata}
(\Kt)_{nm}  &=&   -(n-1) \delta_{n-1,
m}-(n+1) \delta_{n+1, m}\, , \nonumber \\
(\K)_{nm}  &=&   -\sqrt{n(n-1)} \delta_{n-1,
m}-\sqrt{n(n+1)} \delta_{n+1, m}\, .
\een
Since $K_1$ maps twist even to twist odd states,
the associated matrices anticommute with the matrix $C$,
\be
\{ \K, C \} = \{ \Kt, C \} = 0 \,.
\ee
Finally, since $K_1$ is invariant under hermitian conjugation, we have:
\ben
\label{intm1} 
&&  [K_1, v \cdot a^\dagger] =- (\K v) \cdot a^\dagger + 
\sqrt 2 v_1 p \,.
\een

\medskip
It is also convenient to represent vectors of type $w$
($\alpha_n$ basis) in terms of formal power series in a variable $z$,
\be \label{fw}
f_w(z) \equiv \sum_{n=1}^\infty w_n z^n \,.
\ee
With this definition we note that
\be
\label{caction}
f_{Cw} = f_w (-z)\,.
\ee
Using eq.\refb{ecomm} (or directly from  
eq.\refb{kdata}) we see 
that
the operator $K_1 = L_1 + L_{-1}$ on the basis of functions of $z$
is represented by the differential operator:
\be
{\cal K}_1 \equiv -(1+z^2) \frac{d}{dz}\, .
\ee
More specifically,
with the proviso that constant terms obtained after the action
of the differential operator are to be
dropped, we have
\be
f_{\Kt w}(z) = {\cal K}_1 f_w (z)\, .
\ee
It immediately follows from this equation 
that  
\be  \label{eigendiff}
\Kt w^{(\kappa)} = \kappa \, w^{(\kappa)}  \quad \leftrightarrow \quad
{\cal K}_1 f_{w^{(\kappa)}} =
\kappa f_{w^{(\kappa)}} + a\, ,  
\ee
where the constant $a$ is used
to account for the fact that the action of the differential   
operator must be supplemented by removing the constant term.
Therefore the eigenvalue problem for the infinite matrix $\Kt$ can
be studied as the eigenvalue problem for the differential operator
${\cal K}_1$ on the space of formal power series. The differential
equation above is readily integrated to find
\be
\label{sole}
f_{w^{(\kappa)}}(z) =-{1\over \kappa} \exp(-\kappa \tan^{-1}(z)) 
+{1\over \kappa} \equiv \sum_{n=1}^\infty w^{(\kappa)}_n z^n\, ,
\ee
where 
the overall normalization
has been chosen so that $w^{(\kappa)}_1 = 1$. Expanding the above in
powers of $z$ and using \refb{fw} one can read the coefficients
$w^{(\kappa)}_n$ which, because of \refb{eigendiff} provide an eigenvector
of $\Kt$. This shows 
that $\Kt$ has a {\it non-degenerate continuous}
spectrum. We can take $-\infty < \kappa < \infty$, and there is exactly
one eigenvector for each value of $\kappa$. Note also that each
eigenvector of $\Kt$ provides an 
eigenvector $w^{(\kappa)}_n$  
of $\K$ with the same
eigenvalue using the relation 
$v_n^{(\kappa)} = \sqrt{n} w_n^{(\kappa)}$. 
This 
follows from eq.\refb{vw}.

It follows
from
\refb{caction} and \refb{sole} that $Cw^{(\kappa)} = - w^{(-\kappa)}$ and
therefore we can form linear combinations of definite twist
\be  
\label{hl}  
w^{(\kappa)}_\pm \equiv {1\over 2}  \Bigl( w^{(-\kappa)} \mp
w^{(\kappa)}\Bigr)\,,  
\ee
that satisfy
\be
\label{etwist} 
C\, w^{(\kappa)}_\pm  = \pm w^{(\kappa)}_\pm \,.
\ee
The function representation of these
eigenvectors follows from eqs.\refb{hl} and \refb{sole}
\ben
\label{eexp} 
f_{w^{(\kappa)}_-} &=& {1\over \kappa} \sinh(\kappa \tan^{-1}z)
\,, \nonumber \\
f_{w^{(\kappa)}_+} &=& 
{1\over \kappa} \Big(\cosh(\kappa
\tan^{-1}z) - 1) \, .
\een
These definite twist vectors  
are not eigenvectors of $\Kt$ but
are eigenvectors of $\Kt^2$ with eigenvalue $\kappa^2$. For
$\kappa^2>0$ the spectrum of $\Kt^2$ is continuous and doubly
degenerate, having, for each $\kappa^2$ a $C$-even and a $C$-odd
eigenvector. For
$\kappa^2=0$ there are also two eigenvectors. One of them
arises from the $\kappa=0$ eigenvector of $\Kt$. This is obtained
by taking the limit
$\kappa\to 0$ in \refb{sole} and it gives the
$C$-odd eigenvector $w^{(0)} = \tan^{-1}(z)$.
Using eq.\refb{vw} 
we see 
that this 
eigenvector is precisely the $\lambda = -1/3$ eigenvector $v^-$ of
$M$ given in  \refb{eigen33}.  This is no coincidence, 
as we shall 
explain in the next subsection. 
The second zero eigenvector of $\Kt^2$ corresponds to the
function
$(\tan^{-1}(z))^2$. Explicitly it takes the form
\be
\label{ceven}
v_{2k}^+ = {(-1)^{k+1} \over \sqrt{2k}} \Bigl( 1 + {1\over 3} + {1\over
5} + \cdots + {1\over 2k-1} \Bigr)\,, \quad v_{2k-1}^+=0\,, \quad k=1,
2,3
\cdots
\ee
We note, however, that while the norm of $v^-$ diverges logarithmically, the 
norm of $v^+$ has worse divergence.
Furthermore $v^+$ is not an 
eigenvector of $\K$. 
We shall also argue later that truncation of $v^+$ to a given level never 
appears as an eigenvector of the 
level truncated $\K^2$. 
For $\kappa^2 <0$, or imaginary $\kappa$, the spectrum of $\Kt^2$
is still continuous and doubly degenerate. Nevertheless, the norm
of the corresponding eigenvectors seems even more divergent than
the norm of the eigenvectors with real $\kappa$, and we have
seen no evidence of these eigenvectors in our numerical work.
This is of course 
consistent with the fact that $\K$, being a real symmetric matrix, should 
only have real eigenvalues.

\sectiono{Wedge States, $\K$ and $M$} \label{sa4}  
A general wedge state $|N\rangle$ can be expressed as 
\be
\label{wedgeexp}
|N \rangle  = \exp \left(-\frac{1}{2} a^\dagger \cdot (C T_N) \cdot
 a^\dagger\right) |0 \rangle  \equiv \exp \Bigl( - E_N \Bigr) |0\rangle \, ,
\ee
where \cite{0107101}
\be \label{fo}
T_N = \frac{T + (-T)^{N-1}}{1 - (-T)^N} \,.
\ee
$T$ has been defined in eq.\refb{edeft}. 
The matrices $T_N$ are  related
to the Neumann coefficients of the $N-$th complete overlap
string vertex, $V_N = C T_N$.
Important special cases are
$T_\infty = T$  (the sliver)\footnote{This follows from \refb{fo} if 
the eigenvalues of $T$ lie in the range $[-1,0]$, as has been found 
numerically. We shall return to this point later.} 
and $T_3 = M$.

The eigenvectors of $T_N$ are the same for all the matrices
in the family, and the eigenvalues are related according
to the above formula. 
So we can simply focus on $T$ and/or $M$. 
We shall first establish that the eigenvectors of $\K$
are eigenvectors of $M$ and of $M^{12}$ and $M^{21}$.  
Then we find the eigenvalues of $M, M^{12}$,
and $M^{21}$ corresponding to a given
eigenvector.

\subsection{Eigenvectors of $T$ and $M$}   

In this subsection we shall show that  
the eigenvectors of $\K$
are eigenvectors of $M$ and of $M^{12}$ and $M^{21}$.
We will do this in two stages. We first show this is
true for $M$ and all wedge state matrices $T_N$.
Then we turn to the case of the matrices $M^{12}$ and $M^{21}$.

\medskip
We first note that the derivation  
$K_1$ annihilates all wedge states $|N \ra$.  Indeed, $K_1$ annihilates
the identity, which corresponds to $N=1$, and the SL(2,R) vacuum,
which corresponds to $N=2$. Since all higher $N$ wedge states
can be obtained by star multiplication of $N=2$ states we have
$K_1 |N\rangle =0$. We now show that as a consequence of this,
the matrices $T_N$ 
commute with
$\K$.  Indeed, using eq. \refb{wedgeexp} we have   
\be
0= K_1 |N \rangle = K_1  \exp \Bigl( - E_N \Bigr) |0\rangle 
= - [K_1, E_N ] \exp\Bigl( - E_N \Bigr) |0\rangle\, ,  
\ee
since $[K_1, E_N ]$ commutes with $E_N$.  
Using \refb{intm1}, and noting that the momentum operator
kills any wedge state, we have  that the above equation gives 
\be
0= \left( \frac{1}{2}
a^\dagger \cdot \left(\K CT_N + C T_N \K
\right) \cdot a^\dagger \right) |N \rangle =
\left( \frac{1}{2} a^\dagger \cdot  
( C \, [T_N, \K] ) \cdot a^\dagger \right) |N \rangle \,.
\ee
Since the multiplicative factor acting on the wedge states above
consists of creation operators only, the factor itself must vanish
identically.  This implies that 
\be
\label{vcom}
[T_N,  \K] = 0\, .
\ee
Since the spectrum of $\K$ is non-degenerate,  {\it all} eigenvectors
of $\K$ must be  eigenvectors of $T_N$. 
Furthermore, since $T_N$ commutes with $C$, we see from eq.\refb{etwist} 
that $w^{(\pm \kappa)}$ describe degenerate eigenvectors of $T_N$, and 
$w^{(\kappa)}\pm w^{(-\kappa)}$ are simultaneous eigenvectors of $T_N$ and 
$C$.
We should note, however, 
that the relation $[T_N, \K]=0$ holds only for 
infinite 
dimensional matrices $T_N$ and $\K$ and is only appproximate if we 
truncate $T_N$ and $\K$ to finite dimensional matrices.

A similar argument can be used to show that
$\K$ commutes with the matrices $M^{12}$ and $M^{21}$.
The derivation property of $K_1$ implies that
\be
\Bigl(K_1^{(1)} + K_1^{(2)} + K_1^{(3)} \Bigr) |V_3 \rangle = 0 \,,
\ee
where the expression for the vertex was given in \refb{e9} 
and \refb{e9p}. It suffices for the present purposes to work
at zero momentum, and the above equation implies that 
\be
\Bigl(K_1^{(1)} + K_1^{(2)} + K_1^{(3)} \Bigr) 
\, \exp \Bigl( - {1\over 2} \sum_{r,s} a^{(r)\dagger} 
C M^{rs}a^{(s)\dagger} \Bigr) |0 \rangle = 0 \,.
\ee
Since the $K$'s annihilate the vacuum we pick commutators that
give
\be
\Bigl( {1\over 2} \sum_{p,q} a^{(p)\dagger} 
C\, [\K ,  M^{pq}\,] \, a^{(q)\dagger} \Bigr) 
\, \exp \Bigl( - {1\over 2} \sum_{r,s} a^{(r)\dagger} 
C M^{rs}a^{(s)\dagger} \Bigr) |0 \rangle = 0 \,.
\ee
This condition implies that $[K_1, M^{pq}] =0$, as claimed. Once again,
the non-degeneracy of the $\K$ spectrum implies that $\K$ eigenvectors
are eigenvectors of 
$M^{12}$ and $M^{21}$. 

\subsection{Relating the eigenvalues of  $M$ and $\K$ via the $B$ 
matrices}  

While the $\K$ operator helps us determine the eigenvectors of $M$ 
and $T_N$, so far it has not given us information about the corresponding 
eigenvalues as the precise relation between $M$ and $\K$ has not 
been found. In 
this subsection we shall find this relation. We do this
by
introducing a new matrix $B$, much simpler than $M$,
and that shares 
all 
the eigenvectors of $M$.
The relation of $B$ to $T$ (or $M$) is calculable
analytically and thus the relation between their eigenvalues
is fixed. Furthermore the action of $B$ on the eigenvector 
$w^{(\kappa)}_n$ will also be calculable analytically. This in turn, will 
determine the action of $M$ and $T$ on $w^{(\kappa)}_n$.

\medskip
We define $B$ as the leading expansion
of $T_N$, when $N$ is very close to two:
\be
\label{aexp}
T_{2 +\epsilon} = \epsilon \, B + O(\epsilon^2) \,.
\ee
Expanding the formula (\ref{fo}) we find the relation
of $B$ with the sliver matrix $T$,
\be
\label{aft}
B = - \frac{T \ln(-T)}{1-T^2} \,.
\ee
Notice that as $T \to -1$, $B \to -1/2$, and as
$T\to 0$, $B\to 0$.  So the spectrum of $B$ is
expected to lie on the interval $[-1/2, 0]$.

To obtain an expression for
the matrix elements of $B$ we consider in more
detail the wedge state $|
2 +\epsilon\rangle$. We have, on the one hand, 
\be
\label{vire}
|2 + \epsilon \rangle = \exp (\epsilon V_-) |0\rangle =  | 0 \rangle +
\epsilon V_- | 0
\rangle + O(\epsilon^2)
\ee
for an appropriate vector field 
\be \label{evect}  
V_- = \sum_{n=2}^\infty v_n L_{-n}\, .
\ee
To find $V_-$, recall that wedge states $|N \rangle = \exp
\left(\sum_{k=2}^\infty c^{(N)}_k L_{-k} \right)|0\rangle$ are defined
by requiring\cite{0006240}
\be \label{exyy}
\exp \left( \sum_{k=2}^\infty c^{(N)}_k z^{k+1} \partial_z \right) z =
\frac{N}{2}  \tan \left( \frac{2}{N} \arctan(z) \right)
\ee
{}From eqs.\refb{vire}-\refb{exyy} we have $c_k^{(2+\epsilon)}=\epsilon\, 
v_k$.  
Expanding the right hand side of \refb{exyy} for
$N= 2+ \epsilon$ we get 
\be \label{evn}  
\sum_n v_n z^{n+1} = {1\over 2} ( z - (1+z^2) \tan^{-1} z) \, .
\ee
This gives\footnote{This vector field has also been considered
independently by Schnabl\cite{schnabl}.}  
\be
\label{epve} 
V_- =\sum_{n=1}^\infty (-1)^{n} {1 \over {
(2n-1)(2n+1) }   } L_{-2n}  \,.
\ee
On the other hand we also have from \refb{wedgeexp} and
\refb{aexp} that
\be
\label{wedep}
|2+\epsilon \rangle  = \exp \left(-\epsilon \frac{1}{2} a^\dagger \cdot
(C B) \cdot
 a^\dagger\right) |0 \rangle  = |0\rangle -
\epsilon\, \frac{1}{2} a^\dagger \cdot (C B) \cdot
 a^\dagger|0 \rangle\, + O(\epsilon^2).
\ee
Comparing the right hand sides of \refb{vire} and \refb{wedep}, using 
eq.\refb{epve}, and the equation
\be \label{eldefn}  
L_{-m} = {1\over 2}\, \sum_p\, \alpha^\mu_{-m+p} \alpha^\nu_{-p} 
\eta_{\mu\nu} \quad 
\hbox{for} \quad m>0\, ,
\ee
we finally find
\ben \label{bvalue}  
B_{mn} &\equiv& - \frac{(-1)^{\frac{n-m}{2}} \; \sqrt{m \,n}
}{(m+n)^2-1}
\quad {\rm for} \; n+m \; {\rm even} \\
\nonumber
B_{mn} &\equiv& 0 \quad {\rm for} \; n+m \; {\rm odd}
\,.
\een
Note that the matrix $B$ is much simpler than
the matrix
$T$ or $M$.  

\bigskip
Since $T_N$ commutes with $\K$ for every $N$, 
so must $B$. Thus the 
eigenvectors 
$v^{(\kappa)}_n=\sqrt{n} w^{(\kappa)}_n$ must also be eigenvectors of $B$. 
Our goal now is to 
find an expression for the eigenvalues
$\beta(\kappa)$ of $B$
associated to the eigenvectors $v^{(\kappa)}$. 
For this we consider the eigenvalue equation
\be \label{efi1}
\sum_{n\ge 1} B_{mn} v^{(\kappa)}_n = \beta(\kappa) v^{(\kappa)}_m\, .
\ee
Since this relation holds for every $m$ we have, in particular,
\be \label{efi2} 
\beta(\kappa) =\frac{1}{v^{(\kappa)}_1}  \sum_{n\ge 1} B_{1n} v^{(\kappa)}_n 
=  \frac{1}{w^{(\kappa)}_1} \sum_{q=1}^n {(-1)^q\over 2q +1}\,
w^{(\kappa)}_{2q-1} \, .
\ee
If we define
\be \label{efi3}
F(z) =  \sum_{q=1}^n {(-1)^q\over 2q +1}\,\,
w^{(\kappa)}_{2q-1}\, z^{2q +1}\, ,
\ee
then we may rewrite \refb{efi2} as
\be \label{efi4}
\beta(\kappa) = {F(1) \over  w^{(\kappa)}_1} = F(1) \, ,
\ee
since, as seen from eq.\refb{sole}, $w^{(\kappa)}_1=1$.
On the other hand, we have,
\be \label{efi5}
{d F(z) \over d z} = 
\sum_{q=1}^n {(-1)^q} w^{(\kappa)}_{2q-1} z^{2q} 
= {1\over 2}\, i z \, \Big(f_{w^{(\kappa)}}(iz) - 
f_{w^{(\kappa)}}(-iz)\Big)\, ,
\ee
where $f_{w^{(\kappa)}}(z)$ has been defined in eq.\refb{sole}. We can 
easily integrate this equation (with the boundary condition $F(0)=0$) to 
get  
\ben \label{efi6}
\beta(\kappa) = F(1) = - {1\over \kappa} \int_0^1  dz  \, z 
\sin (  \kappa \tanh^{-1}(z)  )  
= -{1\over 2}\, {\kappa\pi/2 \over  
\sinh(\kappa\pi/2)}\, .
\een
This is 
the eigenvalue of the matrix $B$ associated to the eigenvector
$v^{(\kappa)}$. Using
eq.\refb{aft} we  can 
determine the corresponding eigenvalue of $T$ to be:
\be \label{efi7}
\tau(\kappa) = - e^{-|\kappa|\pi/2} \, .
\ee
In deriving \refb{efi7} we have noted that eq.\refb{aft} does not 
determine 
$T$ uniquely for a given $A$, and we 
need some additional 
input. 
This comes from the requirement that the eigenvalue of $T$ lies between 
$-1$ and 0, a fact found in numerical experiments.

Finally, using eq.\refb{fo}, we can determine the eigenvalue $\mu(\kappa)$ 
of $M=T_3$ to be
\be \label{efi8}
\mu(\kappa) = {\tau(\kappa) + (\tau(\kappa))^2 \over 1 + (\tau(\kappa))^3}
= - {1\over 1 + 2\cosh(\kappa\pi/2)}\, .
\ee

This completes the determination of the eigenvectors and eigenvalues of 
the matrix $M$. Furthermore, eq.\refb{efi8} also provides us a simple 
expression for $M$ in terms of the matrix $\K$:
\be \label{emexp}
M = - ( 1 + 2 \cosh(\K \pi / 2) )^{-1}\, .
\ee

\subsection{Diagonalization of $\rho_1$, $\rho_2$, $M^{12}$ and $M^{21}$}

In section \ref{sa1} we defined the two
real symmetric projectors
$\rho_1$ and $\rho_2$ (eqn.~\refb{edefrho}).
Having shown that the eigenvectors of $\K$ are
eigenvectors of $T$ and $M^{rs}$, it follows that they 
must also be eigenvectors of $\rho_1$ and $\rho_2$.  
This implies % can be shown to imply
that the simultaneous eigenstates of $\rho_1$, $\rho_2$
and $M$ are given by the vectors $v^{(\kappa)}_n= \sqrt
n\,w^{(\kappa)}_n$, with $w^{(\kappa)}_n$ defined as in \refb{sole}.
On the other hand $\rho_i$ do not commute with
$C$,  instead a conjugation by $C$ converts $\rho_1$ to
$\rho_2=1-\rho_1$ and vice versa.
Thus the 0 and 1 eigenvalues of $\rho_1$ (or $\rho_2$)
must get exchanged under the action of
$C$.

One could now ask: what is the eigenvalue $\lambda_i(\kappa)$ of 
$\rho_i(\kappa)$ for a given value of $\kappa$? $\lambda_i(\kappa)$ must 
take values 0 or 1. Furthermore we have the relation 
\be \label{eadd}
\lambda_1(\kappa) + \lambda_2(\kappa) = 1\, .
\ee
Also from the twist properties of $\lambda_i(\kappa)$, 
we have  
\be \label{erev}
\lambda_1(-\kappa) = \lambda_2(\kappa) = 1 - \lambda_1(\kappa)\, .
\ee
By continuity in $\kappa$, the only possible choices 
seem to be that $\lambda_1(\kappa)$, for example, be equal to 
one for all positive $\kappa$, or equal to one for all
negative 
$\kappa$. Numerical results show that the 
first possibility is realised:
\be \label{epos1}
\lambda_1(\kappa) =\cases{ 1 \quad \hbox{for} \quad \kappa>0\, ,
\cr  0 \quad \hbox{for} \quad \kappa<0\,. }
\ee    
Thus $\rho_1$ and $\rho_2$ project onto eigenstates of $\K$ with  
positive and negative eigenvalues respectively.

Note that acting on an eigenvector of $M$ with precisely $-1/3$ 
eigenvalue, $\rho_1$ and $\rho_2$ become ill-defined since $(1+T)$ 
vanishes acting on such a state. It is natural to define $\rho_i$ such 
that both $\rho_1$ and $\rho_2$ annihilate this state. Since $\rho_1$ and 
$\rho_2$ project onto the modes of the right- and the left-half of the 
string respectively\cite{0105058,0110204,0105059,0106036}, 
we can interpret
the states  with $\kappa>0$, $\kappa<0$ and $\kappa=0$ as the modes of
the right  half-string, left half-string and the string mid-point
respectively.

Using eqs.\refb{em12m11} and \refb{efi8} one readily shows
that
\be
\label{signam}
\mu^{12} (\kappa)  - \mu^{21} (\kappa) = \pm \sqrt{(1-\mu(\kappa))
(1 + 3\mu(\kappa))}  = \pm
{2
\sinh (\pi\kappa/2)
\over 1 + 2\cosh(\pi\kappa/2)} \, ,
\ee
where $\mu^{12} (\kappa)$ and   $\mu^{21} (\kappa)$ are
the eigenvalues of $M^{12}$ and $M^{21}$ for the eigenvector
$w^{(\kappa)}$. Together with the relation
$\mu^{12} (\kappa)  + \mu^{21} (\kappa) = 1- \mu(\kappa)$ following
from the first equation in \refb{em12m11}, the values of $\mu^{12}
(\kappa)$ and   $\mu^{21} (\kappa)$ would be determined in terms
of $\kappa$ were it 
not for the square root sign ambiguity above. This ambiguity can
be resolved using 
\refb{epos1}.  Indeed, consider $\kappa >0$ in which case 
$\tau (\kappa) = - \exp(-\kappa \pi/2)$ (eqn. \refb{efi7}) and
$\rho_2 w^{(\kappa)} = 0$. Using the explicit form of $\rho_2$
in \refb{edefrho} one can check that $\rho_2 w^{(\kappa)} = 0$
requires choosing the top sign in \refb{signam}.
We thus conclude that
$v^{(\kappa)}$ is  an eigenstate of $M^{12}$ and $M^{21}$ with 
eigenvalues
\ben \label{em12a}
\mu^{12}(\kappa) &=& {1 + \cosh(\pi\kappa/2) + \sinh (\pi\kappa/2) \over
1 +  2 \cosh(\pi\kappa/2)}, \nonumber \\
\mu^{21}(\kappa) &=& {1 +
\cosh(\pi\kappa/2) - 
\sinh (\pi\kappa/2) \over 1 +
2 \cosh(\pi\kappa/2)}\, .
\een

\subsection{String Functionals and the Spectrum}

In this subsection we give the functional interpretation
of the eigenvalue equations we have considered so far.
In this setup one writes the eigenvalue equations as
functional constraints satisfied by the string functionals
associated to wedge states or the sliver. This represents
a generalization of the considerations of Moore and Taylor
\cite{0111069} who interpreted the $C$-odd $\kappa=0$ eigenvector
of the sliver as the existence of a flat direction in the
sliver functional.

To develop this approach in general we define
\be
\label{fint}
T_N \, v^{(\kappa)}_\pm  = \mu_N(\kappa) v^{(\kappa)}_\pm \,, 
\qquad  C v^{(\kappa)}_\pm = \pm v^{(\kappa)}_\pm \,, 
\ee
where $T_N$ is the general wedge state matrix appearing
in \refb{wedgeexp} and \refb{fo},  the vectors 
$v^{(\kappa)}_\pm$ are 
defined in eqs.\refb{evnk}, \refb{elcomb},  
and
$\mu_N(\kappa)$ is the eigenvalue, calculable from 
\refb{fo} and \refb{efi7}.  It follows from \refb{fint}
and \refb{wedgeexp} that
\be
v^{(\kappa)}_\pm \cdot \Bigl(  a  \pm \mu_N(\kappa) \, a^\dagger
\Bigr) |N\rangle = 0 \,, 
\ee
or equivalently
\be
\label{step2}
v^{(\kappa)}_\pm \cdot \Bigl(  (1 \pm \mu_N(\kappa) )
(a+ a^\dagger)   + 
 (1 \mp \mu_N(\kappa) )  \, (a - a^\dagger)
\Bigr) |N\rangle = 0 \,.
\ee
The translation to functional language is effected with
the relations\footnote{These are compatible with the normalization 
convention given in footnote \ref{ff2}.} 
\be  
\hat x_n = {i\over \sqrt{n}} (a_n - a_n^\dagger) \,, \qquad
\hat p_n = -i {\partial \over \partial x_n} = {\sqrt n\over 2}
(a_n + a_n^\dagger) \,,
\ee
which allow us to rewrite \refb{step2} as
\be
\label{step3}  
\sum_{n\geq 1} \Bigl\{   2(1 \pm
\mu_N(\kappa) ) (w^{(\kappa)}_\pm)_n  {\partial \over \partial
x_n}   + (1 \mp \mu_N(\kappa) ) \, n  (w^{(\kappa)}_\pm)_n 
  \,x_n 
\Bigr\} \, \langle X(\sigma) |N\rangle = 0 \,,
\ee
where we used the standard properties
$\langle X(\sigma) | \hat x_n = \langle X(\sigma) | x_n$
and $\langle X(\sigma) | \hat p_n = -i {\partial \over \partial
x_n} \langle X(\sigma) | $ of the position eigenstate. Making use
of the mode expansions 
\be
\label{modeexp}
X (\sigma) = x_0 + \sqrt{2} \sum_{n=1}^\infty  x_n \, \cos
n\sigma\,,\qquad
\pi\, {\delta\over \delta X(\sigma)} = {\partial \over \partial
x_0} + \sqrt{2} \sum_{n=1}^\infty  {\partial \over \partial x_n}
\, \cos n\sigma\, , 
\ee 
we can rewrite the constraint in \refb{step3}   
\be
\label{meq}  
\Bigl\{ 2 (1 \pm \mu_N(\kappa) ) \int_0^\pi d\sigma\,
F^{(\kappa)}_\pm (\sigma)  \pi\,{\delta\over \delta
X(\sigma)}  + (1 \mp \mu_N(\kappa) ) \int_0^\pi  d\sigma
\widetilde F^{(\kappa)}_\pm (\sigma)  \,  X(\sigma)
\Bigr\} 
\, \langle X(\sigma) | N \rangle = 0 \,,
\ee 
where the functions $F$ and $\widetilde F$ are simply related
to the formal functions $f_{w^{(\kappa)}_\pm }(z)$ 
(see \refb{eexp}) representing
the eigenvectors
\ben
\label{meqdef}
F^{(\kappa)}_\pm (\sigma) &=& \sum_{n\geq1}
\,\,\,(w^{(\kappa)}_\pm)_n \,\cos n\sigma \,=\, \Re\Bigl\{
f_{w^{(\kappa)}_\pm} (z)\Bigl|_{z=e^{i\sigma}} \Bigr\}\,,
\nonumber \\
\widetilde F^{(\kappa)}_\pm (\sigma) &=& 
\sum_{n\geq1}
n(w^{(\kappa)}_\pm)_n \cos n\sigma = \Re \Bigl\{ z
{d\over dz} f_{w^{(\kappa)}_\pm} (z)
\Bigl|_{z=e^{i\sigma}}\Bigr\}\,.
\een
Equation \refb{meq}, with the definitions in 
\refb{meqdef},
is the functional constraint associated to the eigenvalue
equation \refb{fint}.  For the particular case of the sliver
$(N\to
\infty)$ we have that $T$ has a $C$-odd eigenvector with
eigenvalue $\mu= -1$ associated to $\kappa=0$. Thus if we 
take the lower sign in \refb{meq} 
the second
term vanishes.  Given this, the constraint simply reduces
to the existence of a flat direction on the sliver functional.
This flat direction is defined by the invariance of the
functional under the variation
$X(\sigma)
\to X(\sigma) +
\epsilon F_-^{(0)}(\sigma)$. Here 
\be
 F_-^{(0)}(\sigma) = \sum_{n\geq1}
\,\,\,(w^{(0)}_-)_n \,\cos n\sigma 
= \Re\Big(\tan^{-1}(e^{i\sigma}) \Big)  
= {\pi\over 4} \cases{+1, \quad 
0\leq \sigma\leq \pi/2 \cr   
-1, \quad \pi/2\leq \sigma\leq \pi\,, } 
\ee
which is the conclusion of \cite{0111069} that the sliver
functional is invariant under opposite rigid displacements
of the left and right halves of the string. Indeed whenever
the second term in \refb{meq} vanishes we have a flat
direction $X(\sigma)
\to X(\sigma) +
\epsilon F_-^{(\kappa)}(\sigma)$. Whenever the first term
vanishes, the  functional $\langle X(\sigma) | N\rangle$  must
contain a delta function of the form $\delta (h(X(\sigma))$
where 
$h(X(\sigma)) =
\int_0^\pi  d\sigma
\widetilde F^{(\kappa)}_\pm (\sigma)  \,  X(\sigma)$. An example
of this case was provided in \cite{0111069}:  a $C$-even
eigenvector of eigenvalue one that is present for the
instantonic sliver, in which case the delta function constraint
requires the midpoint of the string to lie 
at the instanton location.

Given the distribution of eigenvalues
for the wedge states
$|N\rangle$, which lie on the interval $[ -1 + {2\over N} , 0)$,
the eigenvalue $(-1)$ can only be attained for the sliver.
Thus, for all these cases, and for the sliver eigenvalues
different from
$-1$, the constraint is more general. It is a functional
differential constraint on the wave function.  

\sectiono{Spectral Density and  Finite Level Analysis}
\label{sa5}

In the previous sections we have constructed the exact
eigenstates and eigenvectors of the matrices $\K$, $B$, $T$ and $M$. 
The question that we shall 
address in this section is: 
how do these results get modified if we work 
with the truncation of these matrices 
to square $L\times L$ matrices ?  
Clearly, with finite size matrices the eigenvalues will form 
a discrete set.  We explain how this quantization of the 
continuous spectrum arises. We 
then
describe our numerical  results.

\subsection{Quantization condition on the eigenvalues for finite matrices}

We begin by analyzing the matrix $\K$ truncated to a 
square $L\times L$ matrix. By a small abuse in language
we simply call this the level $L$ 
truncation of $\K$,  
and we denote it by 
$\K_L$. 
Since $(\K)_{mn}$ vanishes unless $n=m\pm 1$, 
given 
an eigenstate $v_n^{(\kappa)}$ of the infinite dimensional 
matrix $\K$,  the 
restriction $\vv^{(\kappa)}$ of $v^{(\kappa)}$ to level $L$, defined 
as the $L$ dimensional vector:
\be \label{enum2}
\vv_{n}^{(\kappa)} = v_n^{(\kappa)} \quad \hbox{for $1\le n\le L$}, 
\ee
will 
be an exact eigenstate of $\K_L$ if
\be
\label{quant}
v_{L+1}^{(\kappa)} = 0\,.
\ee
Since $v_n^{(\kappa)}=\sqrt n w_n^{(\kappa)}$ with $w_n^{(\kappa)}$ 
defined through the expansion of eq.\refb{sole}, we see that the 
eigenvalues 
$\kappa$ of $\K_L$ are determined by the equation:
\be \label{enum5}
\kappa^{-1}\, \Bigl(\p_z^{L+1} \exp(-\kappa\tan^{-1} z\Bigr)\Big|_{z=0} 
= 0\, .
\ee
The left hand side of
eq.~\refb{enum5} 
gives a polynomial in $\kappa$ of
degree
$L$, and the $L$ solutions are the eigenvalues of the level $L$ truncation
of $\K$. 
Since $\K_L$ is a real symmetric matrix, all its eigenvalues are 
guaranteed to be real. 
For odd $L$, the polynomial is odd under $\kappa\to -\kappa$ and hence 
always has a solution $\kappa=0$. The corresponding eigenvector is given 
by the restriction of $v_n^-$ defined in eq.\refb{eigen33} to first $L$ 
entries. On the other hand for even $L$ the left hand side of 
eq.\refb{enum5} is an even polynomial in $\kappa$. The constant term in 
this polynomial, proportional to $(\partial_z^{L+1} \tan^{-1}z)|_{z=0}$, 
is 
non-zero for every $L$, and hence there are no zero roots of this 
polynomial. Thus for even $L$ we do not have any eigenvector of $\K_L$ 
with zero eigenvalue.

\medskip
Having explained how the discrete spectrum of the truncated
$\K$ matrix arises, we now turn to the 
eigenvalues of the matrices $B$, $T$ and $M$. 
Let $R$ be any one of the infinite dimensional matrices $B$, $T$ or 
$M$ discussed
earlier, and $u_n(\rho)$ denote its exact 
eigenvector with eigenvalue
$\rho$:
\be \label{enum1}
R_{mn}\, u_n(\rho)  = \rho\, u_m(\rho) \, .  
\ee
It will be convenient to take $u$ to be a simultaneous eigenvector of 
$R$ and $C$ 
(and hence of $\K^2$), given by 
$u_n = v_{\pm n}^{(\kappa)}$,  
with $v_\pm^{(\kappa)}$  
defined as in eqs.\refb{evnk}, \refb{elcomb}.  
Furthermore, let $R_L$ be the 
restriction of $R$ to level $L$, 
and $\uu (\rho)$ be the restriction of the vector $u (\rho)$ to level 
$n\le L$, defined in a manner analogous to eq.\refb{enum2}. 
Then
\be \label{enum3}
\sum_{n=1}^L   
R_{Lmn} \, \uu_{n}(\rho)  = \rho \, \uu_{m}(\rho)  - 
\sum_{n>L} R_{mn} 
\,u_n (\rho) 
\, .
\ee
Now suppose for large $n$, and fixed $m$, the leading contribution 
to $R_{mn}$ has the form:
\be \label{enum4}
R_{mn} \simeq  
 f(m) g(n)\, , \quad  n \gg m\,,
\ee
for some functions $f$ and $g$.
In that case $\uu$ will be an eigenvector of $R_L$ with eigenvalue $\rho$ 
to leading order if
\be \label{enum6}
\sum_{n>L} g(n) \,u_n (\rho) = 0\, .  
\ee
This equation is the approximate quantization condition for the
eigenvalues. 
The specific values of $\rho_i$ 
for which \refb{enum6} is satisfied make the last term in
\refb{enum3} vanish to leading order 
and therefore $(\rho_i , \uu (\rho_i))$
are approximate eigenvalues and eigenvectors of $R_L$.  
Thus the eigenvector associated to a discrete eigenvalue
is simply the level $L$ truncation of the exact  eigenvector of $R$
associated to that eigenvalue. 

\medskip
As already stated above, we choose the vectors $u_n$ to be simultaneous
eigenvectors of $R$ and $C$ (and hence of $\K^2)$. 
The exact 
eigenstates of these infinite dimensional matrices
are given in terms of the expansion coefficients 
of the functions shown in \refb{eexp}.    
The discrete set of $\kappa$'s satisfying eq.\refb{enum4} then gives us 
the approximate eigenstates of $R_L$. The corresponding eigenvalues are 
computed by evaluating $\rho(\kappa)$ given in eqs.\refb{efi6}, 
\refb{efi7} and \refb{efi8} for $R=B$, $T$ and $M$ respectively.
Since the expressions for the function $g(n)$ introduced
in \refb{enum4} in general would 
differ for $B$, $T$ and $M$, at any level 
$L$ the sets of quantized values of $\kappa$ need not agree for these 
different matrices. As a result the corresponding 
eigenvectors will also 
differ from each other, reflecting the fact that the relations between the 
matrices $\K$, $B$, $T$ and $M$, which hold at infinite level, no longer 
hold for the level truncated matrices.

To see examples of the factorization 
property \refb{enum4}, we note that 
if $R$ corresponds to the matrix $B$ defined in 
eq.\refb{bvalue}, 
then for a $C$-odd eigenvector we need only consider 
$m$ and $n$ odd, and we find 
\be \label{enum7}
f(m) = (-1)^{m+1\over 2} \sqrt m, \qquad g(n) = (-1)^{n-1\over 2} 
n^{-3/2}\, .  
\ee
On the other hand, for a $C$-even eigenvector
we need only consider 
$m$ and $n$ even, and
\be \label{enum8}
f(m) = -(-1)^{m\over 2} \sqrt m,  
\qquad g(n) = (-1)^{n\over 2} 
n^{-3/2}\, .
\ee
The factorization described in \refb{enum4}
also holds for $M$ as can be seen, for example, using equation
(4.32) of \cite{0111069}. In the case of $m$ and $n$ odd,
relevant for $C$-odd eigenvectors, one finds $g(n)= (-1)^{{n-1\over 2}}
n^{-{7\over 6}}$, and for $m$ and $n$ even,
relevant for $C$-even eigenvectors, one finds $g(n)= (-1)^{{n\over 2}}
n^{-{7\over 6}}$.

\subsection{Eigenvalue distribution function}

Given the continuous spectrum of eigenvalues, one could study the 
density of eigenvalues.
Since the 
analysis is simplest for the eigenvalues of $\K$, we could first find the 
density 
$\rho_{\K}^L(\kappa)$ of
eigenvalues of $\K_L$, where $\int_{\kappa_0}^{\kappa_1} 
\rho_{\K}^L(\kappa)
d\kappa$ would give the number of eigenvalues of 
$\K_L$ lying between
$\kappa_0$ and $\kappa_1$. We can then take the $L\to\infty$ limit to 
compute 
the asymptotic density $\rho_{\K}(\kappa)$.
This can then be used to compute the density of 
eigenvalues of another matrix (say $M$) via the relation:
\be \label{ereln}
\rho_{M}(\mu) = 2\, \Bigl( {d\mu\over d\kappa}
\Bigr)^{-1} \rho_{\K}(\kappa)\, .
\ee
The factor of two in the above formula comes from the fact that 
two different values of $\kappa$, differing by a sign, give the same 
$\mu$.  

We shall find it more convenient to compute the 
eigenvalue densities 
of 
$C$-odd and $C$-even eigenvectors of $\K_L^2$ separately, and then combine 
the results to find the eigenvalue density 
of $\K_L$. The eigenvalue 
equation for the matrix $\Kt^2$:
\be
\Kt^2 \, w_{(\kappa^2)} = \kappa^2 \,  w_{(\kappa^2)}\,,
\ee
leads to the recursion
relation (we drop the eigenvalue subscript for simplicity) 
\be
w_{n+2}={ {w_n(\kappa^2 -2 n^2)-(n-1)(n-2)
w_{n-2}}\over 
 {(n+1)(n+2)} }. 
\ee
Let us just take $n=2k$ and introduce 
\be
t_k \equiv w_{2k}\,, 
\ee
to consider the $C$-even case.
The equation becomes:
\be \label{ebec}
2(2k+1)(k+1) t_{k+1}= -t_{k}(8 k^2- \kappa^2)-2(2k-1)(k-1) t_{k-1}\,.
\ee
In order to get an eigenvector of ${\wt\K}_L^2$, --
a matrix whose nonvanishing
entries only extend two steps from the diagonal,  --
we need, in analogy to
\refb{quant}, the condition:  
\be \label{ekcond}
t_{[L/2]+1}=0\, ,
\ee
where $[L/2]$ denotes the integral part of $L/2$. Clearly, to solve this 
equation 
for large $L$, we need to find the behaviour of $t_k$ for large $k$. This 
is the problem we shall now address.
We note in passing that 
eq.\refb{ekcond} is never satisfied by the $C$-even, $\kappa^2=0$ 
eigenvector $v^+$ of $\K^2$ given in eq.\refb{ceven}. Thus this 
eigenvector never generates an eigenvector of the level truncated $\K^2$.

Let us introduce $s_k$ as
\be
s_k \equiv k \, t_k \, (-1)^k\, .
\ee
Eq.\refb{ebec} now becomes
\be
(2k+1)  s_{k+1} +(2k-1) \,s_{k-1}= s_{k}(4 k- {\kappa^2\over 2k}) \,.
\ee
We look for 
slowly varying solutions of this equation so that we can regard 
$s_k$ as a continuous function $s(k)$ that does not vary much when 
$k$ changes by one. 
We can then turn this into a differential equation
\be
  s''+{1\over k}s' + {\kappa^2\over 4} {1\over k^2} s = 0 \,, 
\ee
where the leading neglected term in the left hand side is 
${1\over 6} {s'''\over k}$. 
Let us ignore this term for
now -- this will be justified  later. Then
the differential equation is solved by:
\be
s(k) \sim k^\alpha\,, \qquad  \alpha^2 = -{\kappa^2\over 4}\,.
\ee
Therefore, for positive $\kappa^2$ we get
\be
 s \sim  A(\kappa) \cos \left( {{|\kappa|}\over 2} 
\ln k\right)+ B(\kappa) \sin 
\left(  
{{|\kappa|}\over
2} \ln k\right) \,.
\ee
With $s \sim k^\alpha$ the neglected term in the 
differential equation 
\be
{1\over 6} {s''' \over k} \sim {1\over 6} \alpha(\alpha-1)(\alpha-2) {s\over
k^4} \sim {1\over 6} (\alpha-2) {s''\over
k^2} \ll s'' \,, 
\ee
is therefore much smaller than any other term, for any finite $\alpha$
whenever $k$ is large. The same happens for terms with even more
derivatives. Thus this
solution can be trusted. 
 Going back we have 
\ben \label{wform}
w_{2k} &\sim& {(-1)^k\over k} \left[ A(\kappa) 
\cos \left( {{|\kappa|}\over 2} \ln 
k\right)+
B(\kappa) \sin \left( {{|\kappa|}\over
2} \ln k 
\right)\right] \nonumber \\
&\sim&  {(-1)^k\over k} C(\kappa) \sin 
\Bigl( 
{{|\kappa|}\over 
2} \ln
k + \phi(\kappa)\Bigr) \,, 
\een
for some constants $C$ and $\phi$.
(This behaviour is also seen in explicit computations.)

Substituting eq.\refb{wform} into \refb{ekcond} we get
\be \label{fincond}
{1\over 2} \, |\kappa| \, \ln([L/2]+1) + \phi(\kappa) = n \pi \, ,
\ee
for some integer $n$. The difference between successive values of 
$|\kappa|$ satisfying eq.\refb{fincond}, for 
$\ln L >>  |\phi'(\kappa)|$, 
is:
\be \label{egap}
\Delta|\kappa| = {2\pi\over \ln L}\, .
\ee

The analysis of the odd eigenvectors gives a similar equation. Thus we 
have two states in the interval of 
$\Delta|\kappa|=2\pi/\ln L$. 
On the 
other hand, since 
the $\Kt$ (and $\K$) eigenvalues come in pairs with opposite sign, we see 
that if we study the eigenvalues $\kappa$ of the matrix $\K$, they have a 
uniform spacing given by $2\pi/\ln L$ for both positive and negative 
$\kappa$.
This gives
\be \label{rhofor}
\rho_\K^L(\kappa) = {\ln L\over 2\pi} \, .
\ee
Of course this is valid only for finite values of $\kappa$. 
Since the total 
number of eigenvalues for finite $L$ is finite, clearly the distribution 
gets modified for large $|\kappa|$.

The eigenvalue density for the 
large but finite level $L$ truncation
of $M$ now follows by the standard transformation given in
\refb{ereln} with  the derivative 
evaluated using
\refb{efi8}. The result is the one quoted in \refb{mspecden}.

As a consistency check, we now confirm the uniform density
of eigenvalues in $\kappa$ space by an analysis of $M$.
The method is similar to the one used for $\K$ 
and uses Eqs.\refb{enum6}, 
\refb{wform}, and the asymptotic 
behaviour of $g(n)$ ($\sim (-1)^{n/2} n^{-7/6}$) discussed below 
eq.\refb{enum8}. We find 
the following relation for the $C$-even 
eigenvector of $M$:
\be \label{exx1}
\sum_{k=[L/2]+1}^\infty k^{-5/3} \sin 
\Big({|\kappa|\over 2} \ln k + 
\phi(\kappa)\Big) = 0\, .
\ee
Since the summand is a slowly varying function of $k$ for large $k$, we 
can replace the sum by an integral over $k$, and after performing the 
integral, get an equation of the 
form:
\be \label{exx2}
{1\over 2} |\kappa| \ln([L/2]+1) 
+ \chi(\kappa) =n\pi \, ,
\ee
where $\chi(\kappa)$ is another phase factor. Thus for large $L$, the 
eigenvalues of $M$ are uniformly distributed in the $\kappa$ space, with 
the same density as given in \refb{rhofor}, 
as expected. 
Note, however, that since $\chi(\kappa)$ is 
different from $\phi(\kappa)$, the precise values of $\kappa$ appearing in 
the solution of the eigenvalue equation for $M$ differ from that appearing 
in the eigenvalue equation for $\K$.

\subsection{Numerical tests of the spectrum}  

In this subsection we shall present some numerical experiments
we have done on the calculation of eigenvalues and 
eigenvectors of $\K_L$ and $M_L$ at finite level $L$. 
This work confirms the theoretical expectations developed
in the previous sections.

Numerical analysis of the eigenvalues and the eigenvectors
of the matrix $M$ is carried out by constructing these matrices 
following the general procedure given in \cite{gross-jevicki}
and reviewed in section \ref{sa1}, and then 
finding the eigenvalues and eigenvectors of the truncated matrix 
$M_L$ at level $L$. 
It turns out that
even working at finite
level, where the largest eigenvalue is still about 
10\% 
away
from the value $-1/3$, the eigenvectors follow very accurately
the prediction from the $\K^2$ eigenvectors as given in eq.\refb{eexp}. 
Given 
an eigenvector of $M$
and the corresponding eigenvalue determined numerically, we determine 
$\kappa$ using eq.\refb{efi8} and then use this to predict the 
eigenvector using eq.\refb{eexp}.
For example, calculating the spectrum of $M$ to level $L=2048$ we find
that the eigenvalue of largest magnitude is $\mu_0 = -0.310141$, and
it corresponds to the C-odd eigenvector whose first components are
\be
\label{thedata} 
\hskip-6pt w_1 = 1,  \quad w_3 = -0.318455, \quad w_5 = 0.185188, \quad
w_7 = -0.129081, \,\,\,
w_9 = 0.098372\,.
\ee
Note that just like the eigenvalue $\mu_0$ is far from the limiting value 
$-1/3$,  
the eigenvector is also reasonably far from 
the vector $w_-^{(0)}$ associated with the expansion of $\tan^{-1}z$.
Making use of \refb{efi8} we can find the  
associated value  $\kappa(\mu_0)= \pm 0.298782$.
Using \refb{eexp} the C-odd eigenvector of $\Kt^2$ is
\ben
\label{cevena}
f^{(-)}_\kappa (z)= {1\over \kappa} \sinh ( \kappa \tanh^{-1}z)\, .
\een
Expanded in powers of $z$, for $\kappa(\mu_0)$ 
given above, this  
gives
\be
f^{(-)}_{ \kappa(\mu_0)}  (z) = z - 0.318455\ z^3 + 0.185188 z ^5 -
    0.129081  z^7 + 0.098372 z^9 + \cdots
\ee
in remarkable agreement with \refb{thedata}. 
The explanation for this phenomenon has already been discussed in the 
paragraph below 
eq.\refb{enum8}.

\begin{figure}[!ht]
\leavevmode
\begin{center}
\epsfysize=9cm
\epsfbox{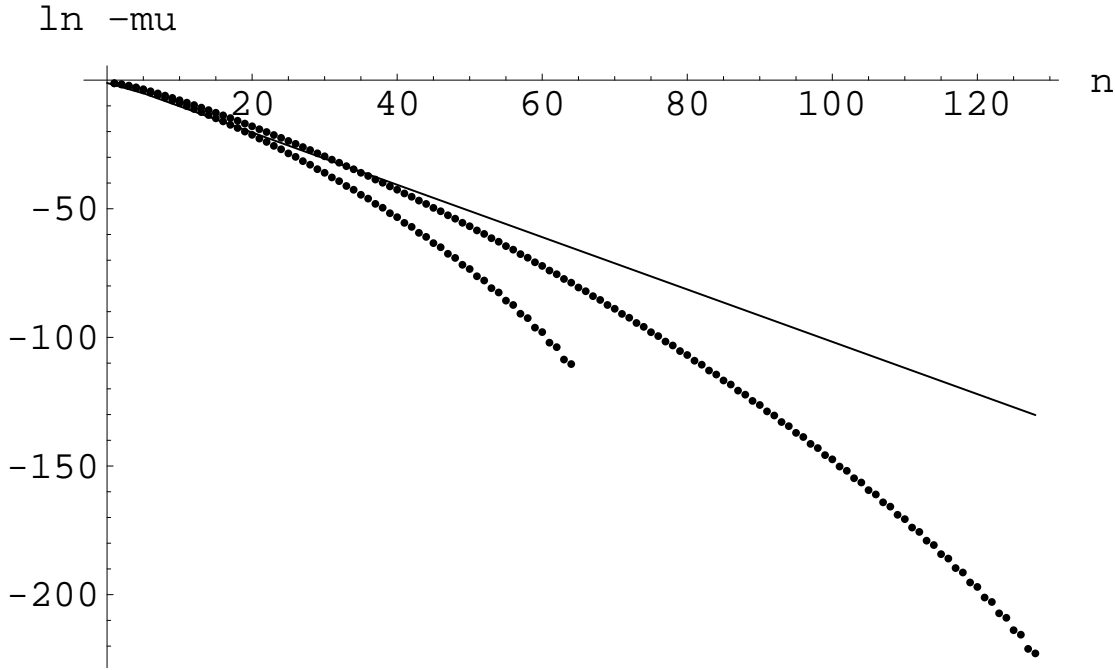}
\end{center}
\caption{This figure shows two plots of eigenvalues $\mu_n$ of the
level truncated matrix $M_L$. In one of them
$L=64$, and in the other $L=128$. On the horizontal
axis we have $n$ referring to the $n$-th eigenvalue
with $n=1$ labelling the smallest eigenvalue (closest to $-1/3$).
On the vertical axis
we show $\ln(-\mu_n)$. Note that the eigenvalues
become small very fast. The solid line shows the predicted curve for 
$L=128$, ignoring corrections of order $1/\ln L$.
}
\label{f1} \end{figure}

The spectrum of eigenvalues of $M$ computed using level 
truncation show some regular pattern. 
Let $\mu_n$ denote the $n$-th eigenvalue of $M$ arranged in 
ascending order so that $n=1$ corresponds to the eigenvalue closest to 
$-1/3$. The eigenvalues corresponding to 
$C$-even
and $C$-odd eigenvectors alternate as their magnitude decrease
monotonically. In fact,  the eigenvalues
go to zero very rapidly and thus one must work with many digits in order
to obtain accurate results.
The eigenvalue spectrum is 
illustrated in Fig.~\ref{f1}, where $\ln(-\mu_n)$
is
plotted against $n$ for the case of level truncations of $M$ with 
$L=64$
and $L=128$. The value of $n$ for any given $\mu$ measures the quantity 
$\int_{-1/3}^\mu \rho^{L}_{M}(\mu') d\mu'$.
As
we can see on the tail ends of the distributions, the pairing of $C$-even
and
$C$-odd eigenvectors emerges as  pairs of dots are seen  to coalesce.
Moreover as we can see, the curve of the level 128 eigenvalues
lies above the curve of the level 64 eigenvalues. 
This is consistent with the emergence of a continuous
spectrum in the $L\to \infty$ level, a fact proven in the previous
sections.

We can compute the predicted answer for $\rho_{M}^L(\mu)$ using 
eqs.\refb{ereln} and \refb{rhofor}. 
This gives 
\be \label{erelint}
n(\mu) =\int_{-1/3}^{\mu}   
\rho^{L}_{M}(\mu') d\mu' = 2\int_0^{\kappa(\mu)} \rho^L_{\K}(\kappa') 
d\kappa' = {\ln L\over \pi} \, \kappa(\mu) \, ,
\ee
in the $L\to\infty$ limit. 
Using eq.\refb{efi8} we can rewrite this as:
\be \label{eintdis}  
n(\mu) 
= 2\, {\ln L\over \pi^2} 
\ln\Big\{{\mu + 1 + 
\sqrt{(1+3\mu)(1-\mu)}\over 2|\mu|}\Big\} \,.
\ee
This predicted curve, computed for $L=128$, has been shown by the solid 
line in Fig.\ref{f1}.

\medskip

\begin{figure}[!ht]
\leavevmode
\begin{center}
\epsfysize=10cm
\epsfbox{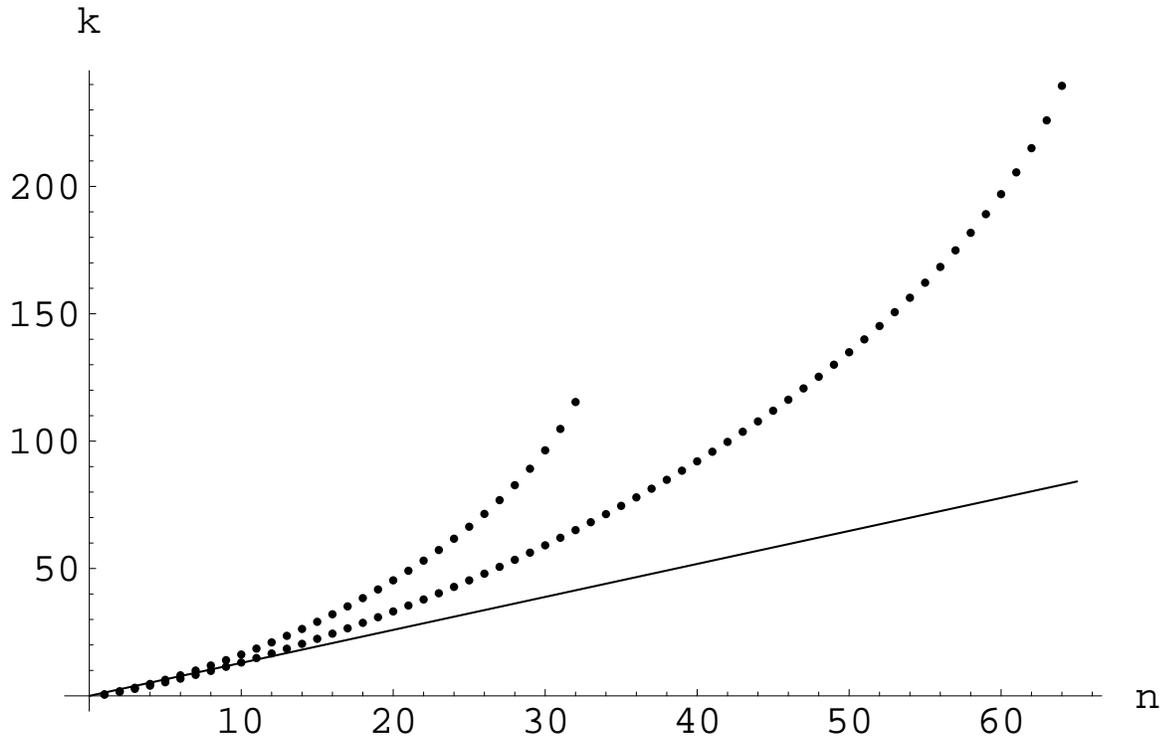}
\end{center}
\caption{This figure shows two plots of the positive
eigenvalues $\kappa_n$ of the
level truncated matrix $\K$. In one of them
$L=64$, and in the other $L=128$. On the horizontal
axis we have $n$ referring to the $n$-th eigenvalue
with $n=1$ labelling the smallest eigenvalue.
On the vertical axis
we show $\kappa_n$. The solid line shows the 
predicted answer for $L=128$ ignoring corrections of order $1/\ln L$.
}
\label{f2} \end{figure}

\bigskip  
Numerical evaluation of the eigenvalues of $\K$ is straightforward using 
eq.\refb{enum5}. Fig.\ref{f2} shows the plot of $\kappa(n)$ vs. $n$, where 
$\kappa(n)$ denotes the $n$-th positive eigenvalue of $\kappa$, with $n=1$ 
representing the lowest positive (or zero) eigenvalue. This data can be 
regarded as a plot of $\int_0^\kappa \rho^L_{\K}(\kappa') d\kappa'$, -- 
the value of $n$ for any given $\kappa$ gives the number 
of $\K$ eigenvalues in the range $[0,\kappa]$.
According to eq.\refb{rhofor}, the predicted answer for this quantity is:
\be \label{kpred}  
n(\kappa) = \int_0^\kappa \rho^L_{\K}
(\kappa') d\kappa'= {\ln L\over 2\pi} \kappa\, .
\ee
We show by the solid line the predicted curve for $L=128$. For small 
$\kappa$ this matches reasonably well the numerical results.

\sectiono{Open questions} \label{sa9} 

In this paper we have diagonalized 
the matrices required for star multiplication
of zero-momentum string functionals.  One immediate
extension that should be contemplated is that of
finding the analogous results for the matrices $M'$ 
introduced in \cite{gross-jevicki,0008252,0102112}  
that
include entries for zero modes. These results would
also determine the spectrum of the matrix defining the
instantonic sliver. Indeed, complete knowledge of the
spectral distributions may allow us to calculate analytically
the ratio
\be
{\det (1-M')^{3/4} \det (1+ 3M')^{1/4}
\over \det (1-M)^{3/4} \det (1+ 3M)^{1/4}}\,
\ee
which enters into the evaluation of the ratios of tensions
of D-branes differing by one dimension
(\cite{0102112}, eqn. (3.10)). Currently the computation
of this ratio can only be done by level expansion, though a BCFT
argument can be used to show that the ratio of tensions must
arise correctly \cite{0105168}.

Another problem of importance is the diagonalization of the Neumann 
coefficients which appear in the computation of the star product in the 
ghost sector.  
This problem, however, is automatically solved given our results, and 
those in ref.\cite{gj2} (eqs.(4.6), (4.7) and (4.11)) relating the matter 
and the ghost Neumann coefficients. In particular, it follows from 
\cite{gj2} that the diagonal Neumann matrix $\wt M^{11}=\wt C \wt V^{11}$ 
appearing in the computation of the star product in the Siegel gauge is 
related to the matrix $M$ analyzed here via the relation:
\be \label{ghost}
\wt M^{11} = - M ( 1+ 2 M)^{-1}\, ,
\ee
up to a simiarity transformation involving scaling of $c_{-n}$ and 
$b_{-n}$ by $\sqrt n$ and $1/\sqrt n$ respectively. Thus the eigenvalues 
and eigenvectors of $\wt M^{11}$ are determined in terms of those of $M$.

In discussing the spectrum of infinite matrices in this
paper we have not restricted ourselves to eigenvectors
with finite norm.  Indeed, under the obvious norm
$\sum_n |v_n^{(\kappa)}|^2 =\infty$. Thus our eigenvectors
are not vectors in a Hilbert space.  This does not seem
to be a problem. The eigenvectors satisfy the eigenvalue
equations in a clear sense: the sums involved converge.
Moreover, our eigenvectors are the ones that indeed
appear to emerge in the finite level analysis. Given the
success of level expansion, this should be taken as strong
evidence that we need to deal with this kind of eigenvectors.
Indeed, our work  
may help understand the proper normalization
condition that should be imposed 
on the eigenvectors. 

At a more basic level, the
analysis in this paper should help build the experience that
will allow a proper understanding of the set
of string fields for which the star algebra is defined.
Currently open string field theory can be viewed as
a formulation of string theory with non-perturbative information,
for example, the existence of 
D-branes. If the proper space
of string field is clearly defined, this would turn open
string field theory into a
{\it non-perturbative definition} of string theory.

\bigskip
\noindent
{\bf Acknowledgements.}  
We would like to thank  D.~Gaiotto, N.~Moeller, M.~Schnabl  
and W.~Taylor for useful discussions.
The work of L.R. was supported in part
by Princeton University
``Dicke Fellowship'' and by NSF grant 9802484.
The  research of A.S. was supported in part by a grant 
from the Eberly College 
of Science of the Penn State University.
The work of  B.Z. was supported in part
by DOE contract \#DE-FC02-94ER40818.

\end{document}